\documentclass{iau}
\usepackage{graphicx}

\title[Nearby Young Stellar Groups] 
{A Pre-Gaia Census of Nearby Stellar Groups}

\author[E.~E.~Mamajek]   
       {Eric E. Mamajek}
\affiliation{Department of Physics \& Astronomy, University of
  Rochester, Rochester, NY, 14627-0171\\email: {\tt
    emamajek@pas.rochester.edu}}

\pubyear{2015}
\volume{314}  
\pagerange{119--126}
\jname{Young Stars \& Planets Near the Sun}
\editors{J. H. Kastner, B. Stelzer, \& S. A. Metchev, eds.}
\begin{document}

\maketitle

\begin{abstract}
The nearest, youngest groups of stars to the Sun provide important
samples of age-dated stars for studying circumstellar disk evolution,
imaged exoplanets, and brown dwarfs.  I briefly comment on the status
of the known stellar groups within 100 pc: $\beta$ Pic, AB Dor, UMa,
Car-Near, Tuc-Hor and $\beta$ Tuc nucleus, Hyades, Col, TW Hya, Car,
Coma Ber, 32 Ori, $\eta$ Cha, and $\chi^1$ For. I also discuss some
poorly characterized groups and ``non-groups.'' Grades for 2015 of
{\it Pass}, {\it Satisfactory}, or {\it Fail} are assigned to the
groups for the purposes of age-dating stars and brown dwarfs. I
speculate that Tuc-Hor could have provided a supernova $\sim$60 pc
away $\sim$2.2 Myr ago which showered the Earth with traces of
$^{60}$Fe-bearing dust.  
\end{abstract}

\firstsection

\section{Introduction}


The stellar groups within 100\,pc offer unique samples to investigate
the results of the star and planet formation process at ages
$\sim$10$^7$-10$^9$ yr.
Members of these groups have provided some of the first and best
examples of imaged dusty debris disks, imaged extrasolar planets, and
young substellar objects.
Given the page limits, I will just summarize some under-appreciated
and new aspects of these groups (and apologize for the lack of
figures).
Recent discussions on these groups can be found in \cite[Zuckerman \&
  Song (2004)]{Zuckerman04}, \cite[Torres et al. (2008)]{Torres08},
\cite[Riedel et al. (2014)]{Riedel14}, \cite[Malo et al.
  (2014)]{Malo14}, and \cite[Gagne et al. (2014)]{Gagne14}.
{\bf Table 1} is a compilation of the stellar clusters and
associations within 100 pc (with best estimates of velocities and
ages).
For those interested in adopting ages to stars based on their
membership to one of these kinematic groups (and assuming the star
exhibits some secondary indicators hinting at coevality with the
group), I assign grades of {\it Pass} (\S2), {\it Satisfactory} (\S3),
or {\it Fail} (\S4).
Table 1 only contains the {\it Pass} and {\it Satisfactory} groups.
%


\begin{table}
\caption{Catalog of Stellar Groups Within 100 pc}
\label{groups_100pc}
\begin{center}
\begin{tabular}{llrrrcclrll}
\hline
\\
Group        & Dist & Ref. & $U$   & $V$   & $W$   & $\sigma_U,\sigma_V,\sigma_W$ & $\sigma_v$ & Ref. & Age & Ref.\\
...          & pc   & ...  & km/s  & km/s  & km/s  & km/s                        & km/s       & ...  & Myr & ...\\
\hline
$\beta$ Pic  & $\sim$15$^a$     & 1 & -10.9 & -16.0 &  -9.2 & 0.3, 0.3, 0.3 & 1.5 & 2 &  23\,$\pm$\,3 & 2\\
AB Dor       & 20.1\,$\pm$\,1.6 & 3 &  -7.6 & -27.3 & -14.9 & 0.4, 1.1, 0.3 & 1.0 & 3 & 150$^{+50}_{-30}$ & 4\\
UMa          & 25.2\,$\pm$\,0.3 & 5 &  14.6 &   1.8 &  -8.6 & 0.4, 0.7, 1.0 & 1.4 & 5 & 530\,$\pm$\,40  & 6\\
Car-Near     & 33\,$\pm$\,1     & 5 & -24.8 & -18.2 &  -2.3 & 0.7, 0.7, 0.4 & 1.3 & 5 & $\sim$200 & 7\\
$\beta$ Tuc  & 43\,$\pm$\,1     & 5 &  -9.6 & -21.6 &  -0.7 & 1.0, 1.3, 0.6 & 1.1 & 5 & 45\,$\pm$\,4 & 4\\
Tuc-Hor      & $\sim$48         & 9 & -10.6 & -21.0 &  -2.1 & 0.2, 0.2, 0.2 & 1.1 & 8 & 45\,$\pm$\,4 & 4\\
Hyades       & 46.5\,$\pm$\,0.5 & 10 & -42.3 & -19.1 &  -1.5 & 0.1, 0.1, 0.2 & 0.3 & 11 & 750\,$\pm$\,150 & 6\\
Columba      & $\sim$50         & 1  & -12.2 & -21.3 &  -5.6 & 1.1, 1.2, 0.9 & ...  & ... & 42\,$\pm$\,5  & 4\\
TW Hya       & 53\,$\pm$\,2     & 12 & -11.2 & -18.2 &  -5.1 & 0.4, 0.4, 0.4 & 0.8 & 12 & 10\,$\pm$\,3 & 4\\
Carina       & $\sim$65         & 1  & -10.5 & -22.4 &  -5.8 & 1.0, 0.6, 0.1 & ...  & ... & 45\,$\pm$\,10 & 4\\
Coma Ber     & 87\,$\pm$\,1     & 10 &  -2.4 &  -5.5 &  -0.6 & 0.1, 0.1, 0.1 & 0.4 & 13 & 560\,$\pm$\,90 & 14\\
32 Ori       & 92\,$\pm$\,2     & 4 & -11.8 & -18.5 &  -8.9 & 0.4, 0.4, 0.3 & $\sim$1 & 5 & 22\,$\pm$\,4 & 4\\
$\eta$ Cha   & 94\,$\pm$\,1     & 15 & -10.2 & -20.7 & -11.2 & 0.2, 0.1, 0.1 & 1.5  & 15 & 11\,$\pm$\,3 & 4\\
$\chi^1$ For & 99\,$\pm$\,6     & 5 & -13.1 & -22.1 & -3.7  & 0.4, 0.5, 1.1 & ...  & 5   & $\sim$50?     & 5\\
\hline
\end{tabular}
\end{center}
{\bf Notes:} Velocities are quoted on the standard Galactic coordinate
system where $U$ is towards the Galactic center, $V$ is towards
Galactic rotation ($\ell$ = 90$^{\circ}$), and $W$ is towards the
north Galactic pole (\cite[e.g. Johnson \& Soderblom
  1987)]{Johnson87}.
$\sigma_U$, $\sigma_V$, $\sigma_W$ are uncertainties in the mean
velocities, not velocity dispersion ($\sigma_v$ is an estimate of the
intrinsic 1D velocity dispersion).
{\bf References and Notes:}
1) Does not have well-defined concentration. Distance is to
centroid estimated by \cite[Malo et al. (2014)]{Malo14}.
2) \cite[Mamajek \& Bell (2014)]{Mamajek14}.
3) \cite[Barenfeld et al. (2013)]{Barenfeld13}. 
4) \cite[Bell, Mamajek, \& Naylor (2015)]{Bell15}, see also Bell (this
volume).
5) This work or Mamajek (unpublished).
%
6) \cite[Brandt \& Huang (2015)]{Brandt15}.  \cite[Brandt \& Huang
  (2015)]{Brandt15} have recently revised the Hyades age upward,
however the -1$\sigma$ uncertainty quoted encapsulates recent younger
($\sim$650 Myr) estimates \cite[e.g. de Bruijne et
  al. (2001)]{deBruijne01}.
7) \cite[Zuckerman et al. (2006)]{Zuckerman06}.
8) \cite[Kraus et al. (2014)]{Kraus14}.
9) mean kinematic distance to 120 Tuc-Hor members from \cite[Kraus et
  al. (2014)]{Kraus14} calculated using UCAC4 proper motions and space
motion from Kraus.
10) \cite[van Leeuwen (2009)]{vanLeeuwen09}.
11) \cite[de Bruijne et al.(2001)]{deBruijne01}.
12) Mean distance and velocity using astrometry from \cite[Ducourant
  et al. (2014)]{Ducourant14}, but omitting interlopers TWA 14, 15,
19, \& 22. Velocity agrees well with \cite[Weinberger et
  al. (2013)]{Weinberger13}. Velocity dispersion from \cite[Ducourant
  et al. (2014)]{Ducourant14} and \cite[Mamajek (2005)]{Mamajek05}.
13) Calculated using astrometry from \cite[van Leeuwen
(2009)]{vanLeeuwen09} and radial velocity from \cite[Mermilliod et
    al. (2009)]{Mermilliod09}, and velocity dispersion from
  \cite[Mermilliod et al. (2009)]{Mermilliod09}.
14) \cite[Silaj \& Landstreet (2014)]{Silaj14}.
15) \cite[Murphy et al. (2013)]{Murphy13}.
%
%
\end{table}

\section{Physical Groups (Grade: Pass)}

The {\bf Ursa Major}, {\bf Hyades}, {\bf Coma Ber}, and {\bf $\eta$
  Cha} groups are clearly real {\it clusters}. The young ($\sim$10
Myr) $\eta$ Cha cluster's density is roughly $\sim$30
M$_{\odot}$\,pc$^{-3}$ - the densest of any cluster within 100 pc -
while the older clusters ($\sim$0.5 Gyr) are lower density
($\sim$0.3-3 M$_{\odot}$\,pc$^{-3}$), but all exceed the local disk
density ($\sim$0.1 M$_{\odot}$\,pc$^{-3}$).  The $\sim$10 Myr-old {\bf
  TW Hya} {\it association} is a well-characterized group of $\sim$3
dozen stars \cite[(Kastner et al. 1997, Mamajek 2005, Weinberger et
  al. 2013, Ducourant et al. 2014)]{Kastner97, Mamajek05,
  Weinberger13, Ducourant14}, and will not be discussed further.  The
{\bf $\beta$ Pic} {\it association} was recently reviewed by
\cite[Mamajek \& Bell (2014, and references therein)]{Mamajek14}. Ages
of $<$20 Myr can be discounted as the 6 A-type members plus the F0
member 51 Eri are all on the ZAMS. For the ``classic'' age of 12 Myr
adopted for $\beta$ Pic for most of the past decade, {\it all}
isochrones would predict that late A- and early F-type members should
be very much pre-MS (alas, they are not).  MS turn-on ages and Li
depletion boundary ages appear to be in agreement with a mean age of
23\,$\pm$\,3 Myr.


Table 1 lists the mean distance to the {\bf AB Dor} {\it nucleus}
based on revised Hipparcos parallaxes (\cite[van Leeuwen
  2007]{vanLeeuwen07}) for nuclear members listed by \cite[Zuckerman et
  al. (2004)]{Zuckerman04}.
Omitting HIP 26369 due to large parallax uncertainty, the mean
distance to the rest of the nuclear members is 20.1$^{+1.7}_{-1.4}$
pc, with intrinsic 1$\sigma$ dispersion of $\pm$4 pc.  \cite[Barenfeld
  et al. (2013)]{Barenfeld13} showed that the group must be $>$120 Myr
old, and the new analysis by \cite[Bell et al. (2015)]{Bell15}
estimates an isochronal age of roughly 150$^{+50}_{-30}$ Myr. {\it
  There is no astrophysical support for ages as young as $\sim$50-100
  Myr that are commonly cited}. While the nuclear stars show
remarkably coherent motions (1D dispersion $\sim$1 km\,s$^{-1}$), and
are obviously concentrated, the status of the rest of the AB Dor
membership (i.e. the ``{\it stream}'') is unfortunately murkier
(grade: S/F?).  A spectroscopic analysis by \cite[Barenfeld et
  al. (2013)]{Barenfeld13} showed that only roughly half of purported
AB Dor members sampled outside the nucleus were consistent with being
co-chemical (i.e. possibly co-natal), hence the AB Dor group may share
motions with other young stars of similar ages from different
birthsites.


\cite[Zuckerman et al. (2006)]{Zuckerman06} discovered a nearby
$\sim$200 Myr-old group dubbed {\bf Car-Near}. The mean distance to
the Zuckerman nuclear members
using revised Hipparcos astrometry is 32.7\,$\pm$\,1.2 pc, with median
RV = +17.5\,$\pm$\,0.8 km\,s$^{-1}$.  The intrinsic velocity
dispersion is only 1.3\,$\pm$\,0.5 km\,s$^{-1}$. The group is puny
($\sim$8 M$_{\odot}$), but its inferred density is just below the
local disk density. No objects are hotter than F1, and a thorough
search of the Hipparcos catalog finds no plausible B/A members.


\cite[Kraus et al. (2014)]{Kraus14} have shown the {\bf Tuc-Hor} group
to be a much larger entity ($>$10$^2$ stars) than previously
appreciated.
Projections of the total stellar population based on the members found
so far range from $\sim$200-400 (A. Kraus, priv. comm., \cite[Gagne et
  al. 2014]{Gagne14}, and calculation by author).
The distribution of stars in Tuc-Hor appears somewhat filamentary and
sheet-like over tens of parsecs - even at age $\sim$40 Myr.
While the kinematic data are consistent with a velocity dispersion of
only $\sim$1 km\,s$^{-1}$, the small extent of the group in $Z$
suggests a dispersion of $<$0.2 km\,s$^{-1}$, analogous to that seen
for small scale structures in the Taurus clouds.
Combining the membership lists of \cite[Malo et al. (2014)]{Malo14}
and \cite[Kraus et al. (2014)]{Kraus14}, one starts to see
substructure in Tuc-Hor (\cite[see also Fig. 4 of Zuckerman et
  al. 2001]{Zuckerman01}).
Tuc-Hor is draped across the southern sky, with several ill-defined
clumps which may constitute subgroups:
1) a small clump of members in Pavo associated with the massive star
Peacock ($\alpha$ Pav; B2.5; $\sim$6 M$_{\odot}$; $\alpha, \delta$
$\simeq$ 306$^{\circ}$, -57$^{\circ}$; $\sim$55 pc);
2) another small clump in Indus associated with HR 8352 (HIP 108195;
F1; $\alpha$, $\delta$ $\simeq$ 329$^{\circ}$, -62$^{\circ}$; $\sim$46
pc);
3) another small clump associated with DS Tuc (HIP
116748; G5; $\alpha, \delta$ $\simeq$ 355$^{\circ}$, -69$^{\circ}$;
$\sim$46 pc);
4) the original {\bf $\beta$ Tuc} nucleus (discovered by
\cite[Zuckerman \& Webb 2000]{Zuckerman00}, listed separately in Table
1; $\alpha, \delta$ $\simeq$ 8$^{\circ}$, -63$^{\circ}$; $\sim$43 pc);
5) to the heart of the {\bf Horologium} subgroup centered on Achernar
(B3; $\alpha$, $\delta$ $\simeq$ 24$^{\circ}$, -57$^{\circ}$; $\sim$43
pc)\footnote{The $\sim$8.7 M$_{\odot}$ Achernar binary has {\it 7(!)}
  K7-M5 Tuc-Hor members from \cite[Kraus et al. (2014)]{Kraus14}
  projected within its estimated tidal radius 2.8 pc $\simeq$
  $\sim$3$^{\circ}$.7: 2MASS J01344601-5707564, J01375879-5645447,
  J01504543-5716488, J01380311-5904042, J01505688-5844032,
  J01521830-5950168, J01275875-6032243.  They display a clear
  Li-depletion boundary between $\sim$M4-M4.5 which could be used to
  independently age-date Achernar.};
6) a small clump centered on $\eta$ Hor (A6; $\alpha$, $\delta$ $\simeq$
39$^{\circ}$, -52$^{\circ}$.5; $\sim$46 pc); and
7) a small, diffuse clump near $\epsilon$ Hyi (B9; $\alpha$, $\delta$
$\simeq$ 40$^{\circ}$, -70$^{\circ}$; $\sim$47 pc).
Tuc-Hor does not appear so much as one large group as an ensemble of
evaporating subgroups, with many of the low-mass members in the
immediate vicinity of more massive stars (indeed, likely within the
tidal radii, constituting unstable ``trapezia'').
Precise astrometry and further characterization of the membership of
the Tuc-Hor complex with Gaia should yield a remarkable picture of a
dynamical ``missing link'' between star-forming regions and the field
population.

Given the number of 2-8 M$_{\odot}$ Tuc-Hor members known ($\sim$7), a
Salpeter IMF would predict $\sim$1 star with mass $>$8
M$_{\odot}$. Tuc-Hor may have eked out forming a star hotter than B2
which would have undergone supernova in the recent past. {\it It is
  possible that Tuc-Hor, and not Sco-Cen, was responsible for the
  supernova which produced the 2.2 Myr-old $^{60}$Fe signal in sea
  floor ferromanganese crusts, and contributed to sweeping out the
  Local Bubble} (\cite[e.g. Fry et al. 2015]{Fry15}). Tucana was only
slightly further away ($\sim$60 pc) from the Sun 2.2 Myr ago, well
positioned to provide a $\sim$8 M$_{\odot}$ supernova close enough to
Earth for its $^{60}$Fe-enriched dust to pollute the Earth.

\section{Likely Physical Groups; More Work Needed (Grade: Satisfactory)}

The {\bf Carina} and {\bf Columba} groups are reported to be of
similar ages and kinematics to Tuc-Hor (\cite[Torres et
  al. 2008]{Torres08}). \cite[Bell et al. (2015)]{Bell15} estimate new
consistent isochronal ages of 45$^{+11}_{-7}$ and 42$^{+6}_{-4}$ Myr
for Car and Col, respectively. I have tentatively included these in
Table 1, however they clearly require further study.

{\bf 32 Ori:} This is a new group from \cite[Mamajek
  (2007)]{Mamajek07} that will be discussed further in \cite[Bell et
  al. (2015)]{Bell15} and Shvonski et al. (in prep.). Bell et al. was
able to find a consistent age for a preliminary membership list
\footnote{32 Ori,
  HD 35656,
  HD 35714,
  HD 36338,
  HD 35499,
  HD 35695,
  HD 245059,
  2MASS J05200029+0613036,
  2MASS J05203182+0616115,
  2MASS J05234246+0651581,
  V1874 Ori,
  2MASS J05253253+0625336,
2MASS J05194398+0535021.  Group is centered near $\alpha$,
  $\delta$ $\simeq$ 82$^{\circ}$, +6$^{\circ}$, with proper motion
  $\mu_{\alpha}, \mu_{\delta}$ $\simeq$ +9, -35 mas\,yr$^{-1}$ and
  mean radial velocity +18.5\,$\pm$\,0.4 km\,s$^{-1}$.}, of
22\,$\pm$\,4 Myr.


The {\bf Alessi 13 ($\chi^1$ For)} cluster (\cite[Dias et
  al. 2002]{Dias02}) appears to consists of at least a dozen stars
with masses in the range 0.9-2.4 M$_{\odot}$ (total mass $\sim$18
M$_{\odot}$) - anchored by the A1 stars $\chi^1$ For and $\chi^3$ For
A ($\alpha, \delta$ $\simeq$ 51$^{\circ}$, -36$^{\circ}$;
$\mu_{\alpha}$, $\mu_{\delta}$ $\simeq$ +37, -4 mas\,yr$^{-1}$).
Within a volume of $\sim$100 pc$^3$, these stars have a density of
$\sim$0.2 M$_{\odot}$\,pc$^{-3}$ - roughly double the local disk
density.
\cite[Kharchenko et al. (2013)]{Kharchenko13} estimated an isochronal
age of $\sim$525 Myr, but the MS turn-off is ill-defined.
However, the X-ray emitting stars\footnote{TYC 7027-715-1, TYC
  7027-852-1, TYC 7026-185-1.} {\it all} appear to have saturated
emission (log(L$_X$/L$_{bol}$ $\simeq$ -3.4), more suggestive of an
age more like $\sim$10$^{7.5}$ yr.
Indeed, Alessi 13's velocity and position are suggestive that it may
be part of the Tuc-Hor/Col/Car complex, which itself may be related to
the Cas-Tau association.

\section{Unphysical Groups or Streams (Grade: Fail)}

While there are convincing cases of $\sim$40 Myr-old stars in {\bf
  Argus} membership lists, \cite[Bell et al. (2015)]{Bell15} were
unable to assign an ambiguous isochronal age to the group. The scatter
in HR diagram positions is highly suggestive that Argus is either
seriously contaminated by interlopers, and/or may not constitute a
coeval group (i.e. is a {\it stream}).


\cite[Zuckerman et al. (2013)]{Zuckerman13} proposed a new group of
$\sim$14 systems dubbed {\bf Oct-Near} with age $\sim$30-100 Myr. The
group has a wide range in velocity, most importantly V and W ($>$5
km\,s$^{-1}$). The members do not appear to clump on the sky.  The
uniqueness of the velocities of these young stars is pointed out by
Zuckerman et al., however it probably warrants {\it stream} status at
this point, and further investigation is needed.

{\bf Her-Lyr} was defined by \cite[Gaidos (1998)]{Gaidos98} and
\cite[Fuhrmann (2004)]{Fuhrmann04}, with more recent adjustments by
\cite[Lopez-Santiago et al. (2006)]{LopezSantiago06} and
\cite[Eisenbeiss et al. (2013)]{Eisenbeiss13}.
Fuhrmann has demonstrated that there is a clump in velocity space of
young-ish stars within 25 pc, however given the large velocity
dispersion (variously quoted at $\sim$3-4 km/s), the stars in most of
these lists would not stay within each other's vicinity for very long.
\cite[Lopez-Santiago et al. (2006)]{LopezSantiago06} said their
Her-Lyr members were {\it ``chosen by their kinematics assuming a
  total dispersion of $\pm$6 km\,s$^{-1}$ in $U$ and $V$,
  respectively... [t]he value of the dispersion has been chosen equal
  to that of the $\sim$200 Myr old Castor MG...  coeval with the
  Hercules-Lyra Association. No restriction in the $W$ component has
  been imposed in this first selection.''}
This is a recipe for selecting stars with a wide range of ages and
birthsites (and whose sample mean age is likely to be of limited
utility).
There is remarkably little continuity in the published Her-Lyr
membership lists.
Only three stars have withstood the scrutiny of Fuhrmann,
Lopez-Santiago et al., and Eisenbeiss et al. as Her-Lyr ``members'':
HD 10008, HD 166, and HD 206860, with the latter two being the sole
surviving members from the original Gaidos study!
\cite[Eisenbeiss et al. (2013)]{Eisenbeiss13} does show that the
gyrochronology ages for these three stars are somewhat clustered (231,
315, 296 Myr, respectively).
Extrapolation of the mass function for Her-Lyr by \cite[Eisenbeiss et
  al. (2013)]{Eisenbeiss13} predicts a population of $\sim$25
$\sim$200-300 Myr Her-Lyr M dwarf members within 25 pc.
The spectroscopic and kinematic survey of local ($<$25 pc)
X-ray-bright M dwarfs by \cite[Shkolnik et al. (2012)]{Shkolnik12} was
designed to discover just such young M dwarfs.
Their survey yielded only a {\it single} candidate Her-Lyr M dwarf.
Thus far, Her-Lyr may constitute a {\it stream}, but I concur with
\cite[Brandt et al. (2014)]{Brandt14} that membership to Her-Lyr
should not be used for age-dating stars.


\cite[Mamajek et al. (2013)]{Mamajek13} and \cite[Zuckerman et
  al. (2013)]{Zuckerman13} independently argued that the {\bf Castor
  Moving Group} is unphysical.  Purported Castor ``members'' have a
wide velocity dispersion ($\sim$3-6 km/s). The velocities of the key
members (e.g. Fomalhaut, Vega, etc.)  are sufficiently
well-constrained that one can confidently conclude that they were not
near each other as recently as 10 Myr ago, let alone hundreds of Myr
ago.  The group should probably be considered the {\it Castor Stream}.
The {\bf IC 2391 Supercluster} and {\bf Local Association} are streams
that suffer from the same problems as Her-Lyr and Castor. Membership
to these groups and adoption of group ages is unhelpful for age-dating
stars.


The nearest Cepheid is the famous F8 supergiant {\bf Polaris}, which
is at least a triple system along with two F dwarfs.
\cite[Turner (2004)]{Turner04} proposed that Polaris belonged to a
larger group\footnote{Including HD 7283, 40335, 45919, 52908, 103435,
  108862, \& 118285.}, and I've included it here as there are a range
of distance estimates between $\sim$90-130 pc (\cite[Turner
  2009]{Turner09}).
However, the recent trig parallax from \cite[van Leeuwen
  (2007)]{vanLeeuwen07} places it at $\sim$129 pc.
\cite[Turner (2009)]{Turner09} presented a CMD for Polaris's
neighbors, but no kinematic analysis.
If one adopts the Polaris space motion of \cite[Wielen et
  al. (2000)]{Wielen00} ($U, V, W$) = (-14.2\,$\pm$\,1.2,
-28.0\,$\pm$\,0.8, -5.4\,$\pm$\,1.0) km\,s$^{-1}$, and convert it to a
convergent point solution of $\alpha$, $\delta$ = 107$^{\circ}$,
-31$^{\circ}$, $S_{tot}$ = 31.9 km\,s$^{-1}$, one finds that Turner's
stars are clearly not sharing motion (peculiar velocities of typically
$\sim$5-10 km/s; too large to stay near Polaris over tens of Myr.).
I have seen no evidence for a ``Polaris cluster''.


\cite[Chereul et al. (1999)]{Chereul99} reported the
discovery of three new ``loose clusters'' in a Hipparcos study of
density-volume inhomogeneities among nearby ($d$ $<$ 125 pc) A-F type
stars.
The closest of these ($d$ $\simeq$ 89 pc), was dubbed ``Pegasus 2'' by
Chereul et al., but appears as ``{\bf Chereul 3}'' in
WEBDA\footnote{http://www.univie.ac.at/webda/} and \cite[Dias et
  al. (2002)]{Dias02}.
WEBDA currently lists Chereul 3 as the 3rd closest {\it open cluster},
while \cite[Dias et al. (2002)]{Dias02} call it a {\it moving group},
and exclude it from their open cluster catalog.
Despite their proximity, the Chereul cluster candidates have gone
largely unstudied since their discovery.
The memberships for these groups were not listed in \cite[Chereul et
  al. (1999)]{Chereul99}, however they were kindly provided to the
author by E. Chereul (priv. comm.).
While Fig. 10 of \cite[Chereul et al. 1999]{Chereul99} shows 8
members, the list from Chereul (priv. comm.)  contained 6 Hipparcos
systems\footnote{Chereul 3 members: HIP 103652, 104338, 105902,
  106488, 106783, 107120.  One of these is resolved into two
  components by Hipparcos (HIP 103652) and one has an obvious wide
  separation companion (HIP 106781 is 39'' from HIP 106783).  Two
  A/F-type stars missing from Chereul's list which appear in Fig. 10
  of \cite[Chereul et al. 1999]{Chereul99} are HIP 104430 and
  104616.}.
The revised Hipparcos parallaxes for these 6 stars indicate that they
are at a range of distances (roughly $\pm$20 pc rms about a mean
distance of 95 pc).
Unfortunately, their radial velocities in the compiled catalog of
\cite[Gontcharov et al. (2006)]{Gontcharov06} have a large scatter
(range: -38 to +16 km/s).
Attempts to determine a robust convergent point solution for this
sample which results in reasonable agreement between observed and
predicted radial velocities, and trig and kinematic parallaxes, were
unsuccessful.
The HR diagram positions of the Chereul 3 stars
are consistent with A/F-type main sequence stars drawn from a wide
range of ages (0.4-2.5 Gyr).
HIP 104338 and 104616 (54' separation) could comprise a wide physical
pair (similar RVs and astrometry), but there is little to suggest any
relation between the other purported members.
I conclude that Chereul 3 is likely to be unphysical.


{\bf Chereul 2:} \cite[Chereul et al. (1999)]{Chereul99} also reported
a $\sim$800 Myr-old group dubbed ``Pegasus 1'', but listed as
``Chereul 2'' by WEBDA and \cite[Dias et al. (2002)]{Dias02}.
E. Chereul (priv. comm.) listed 14 candidate members\footnote{Chereul
  2 members: HIP 102299, 103261, 103652, 103813, 104771, 104884,
  105478, 105608, 106362, 107585, 108389, 108439, 108441, 109349, \&
  110465.}, for which I estimate mean distance 91 pc and E(B-V) =
0.04.
However the parallaxes are consistent with intrinsic spread $\pm$15 pc
and the radial velocities range from -28.6 to +14.4 km\,s$^{-1}$,
with no obviously clumping.
The HR diagram positions are consistent with isochronal ages between
the ZAMS and $\sim$10$^{9.4}$ yr, hence both the color-mag and
kinematic data are consistent with Chereul 2 being unphysical.


\cite[Latyshev (1977)]{Latyshev77} reported a candidate nearby open
cluster in UMa (\cite[dubbed {\bf Latyshev 2} by Archinal \& Hynes
  2003]{Archinal03}) comprised of 7 A-type stars of similar proper
motions and magnitudes\footnote{SAO 28803, 28866, 28885, 28928, 28843,
  28868, and 28891. SAO 28868 appears to be a typo in \cite[Latyshev
    (1977)]{Latyshev77}, as the star's properties clearly correspond
  to SAO 28862 instead.}.
With modern astrometry, it is clear that these stars have a wide range
of space velocities (8 to 25 km\,s$^{-1}$), and no consistent
convergent point solution can be found which explains the observed
range of radial velocities (-19 to -3 km\,s$^{-1}$).
The mean distance to the Hipparcos entries is 110 pc, but the
kinematics indicate it is unphysical.


\end{document}